%% file: main.tex
\RequirePackage{lineno}
\documentclass[aps,prl,twocolumn,showpacs,amsmath,amssymb]{revtex4-1}
\usepackage{overpic}
\usepackage{graphicx}
\usepackage{epsfig}
\usepackage{dcolumn}
\usepackage{bm}
\usepackage{rotating}
\usepackage{multirow}
\usepackage{subfigure}
\usepackage{hhline}
\usepackage{amssymb}
\usepackage{lineno}
\usepackage{hyperref}
\usepackage{mathrsfs}
\usepackage{verbatim}
\usepackage{amsmath}
\usepackage{amssymb}
\usepackage{amsfonts}
\usepackage{amsbsy}
\hypersetup{colorlinks,linkcolor=blue,anchorcolor=blue,citecolor=blue}
\include{def-com}

\let\oldequation\equation
\let\oldendequation\endequation

\renewenvironment{equation}
  {\linenomathNonumbers\oldequation}
  {\oldendequation\endlinenomath}

\begin{document}

\title{Stringent test of $CP$ symmetry in $\Sigma^+$ hyperon decays}

\author{
\begin{small}
\begin{center}
M.~Ablikim$^{1}$, M.~N.~Achasov$^{4,c}$, P.~Adlarson$^{76}$, X.~C.~Ai$^{81}$, R.~Aliberti$^{35}$, A.~Amoroso$^{75A,75C}$, Q.~An$^{72,58,a}$, Y.~Bai$^{57}$, O.~Bakina$^{36}$, Y.~Ban$^{46,h}$
% ...
\\
\vspace{0.2cm}
(BESIII Collaboration)\\
\vspace{0.2cm} {\it
$^{1}$ Institute of High Energy Physics, Beijing 100049, People's Republic of China\\
% ...
$^{4}$ Budker Institute of Nuclear Physics SB RAS (BINP), Novosibirsk 630090, Russia\\
% ...
$^{35}$ Johannes Gutenberg University of Mainz, Johann-Joachim-Becher-Weg 45, D-55099 Mainz, Germany\\
$^{36}$ Joint Institute for Nuclear Research, 141980 Dubna, Moscow region, Russia\\
% ...
$^{46}$ Peking University, Beijing 100871, People's Republic of China\\
% ...
$^{57}$ Southeast University, Nanjing 211100, People's Republic of China\\
$^{58}$ State Key Laboratory of Particle Detection and Electronics, Beijing 100049, Hefei 230026, People's Republic of China\\
% ...
$^{72}$ University of Science and Technology of China, Hefei 230026, People's Republic of China\\
% ...
$^{75}$ University of Turin and INFN, (A)University of Turin, I-10125, Turin, Italy; (B)University of Eastern Piedmont, I-15121, Alessandria, Italy; (C)INFN, I-10125, Turin, Italy\\
$^{76}$ Uppsala University, Box 516, SE-75120 Uppsala, Sweden\\
% ...
$^{81}$ Zhengzhou University, Zhengzhou 450001, People's Republic of China\\
\vspace{0.2cm}
$^{a}$ Deceased\\
% ...
$^{c}$ Also at the Novosibirsk State University, Novosibirsk, 630090, Russia\\
% ...
$^{h}$ Also at State Key Laboratory of Nuclear Physics and Technology, Peking University, Beijing 100871, People's Republic of China\\
% ...
}
\end{center}
\vspace{0.4cm}
\end{small}
}

\begin{abstract}
The non-leptonic two-body weak decays $\Sigma^{+} \to p \pi^{0}$ and $\bar{\Sigma}^{-} \to \bar{p} \pi^{0}$ are investigated, utilizing $(1.0087\pm0.0044)\times10^{10}$ $J/\psi$ events and $(2.7124\pm0.0143)\times10^{9}$ $\psi(3686)$ events collected by BESIII experiment.
  The precision of the weak-decay parameters for the decays $\Sigma^{+} \to p \pi^{0}$ ($\alpha_{0}$) and $\bar{\Sigma}^{-} \to \bar{p} \pi^{0}$ ($\bar{\alpha}_{0}$) is improved by a factor of three compared to the previous world average.
Furthermore, the quantum-entangled $\Sigma^{+}\bar{\Sigma}^{-}$ system enables the most precise test of $CP$ symmetry for the decay $\Sigma^+\to p\pi^0$, through the asymmetry observable $A_{CP}=(\alpha_{0}+\bar{\alpha}_{0})/(\alpha_{0}-\bar{\alpha}_{0})$ that is measured to be $-0.0118\pm0.0083_{\rm stat}\pm0.0028_{\rm syst}$.
Assuming $CP$ conservation, the average decay parameter  is determined to be ${\left< \alpha_{\rm 0}\right>} = (\alpha_0-\bar\alpha_0)/2=-0.9869\pm0.0011_{\rm stat}\pm0.0016_{\rm syst}$, which is the most precise measurement of the asymmetry decay parameters in baryon sectors.
The angular dependence of the ratio of the polarization of the $\Sigma^+$  in both $J/\psi$ and $\psi(3686)$  decays is studied for the first time. 
\end{abstract}

\maketitle
\oddsidemargin -0.2cm
\evensidemargin -0.2cm

Violation of charge conjugation and parity ($CP$) symmetry is one of the three fundamental conditions to explain the abundance of matter with respect to antimatter in the universe~\cite{Sakharov:1967dj}. 
$CP$ violation has been observed in decays of $K$, $B$ and $D$ mesons and is described theoretically in the Standard Model (SM) of particle physics through the Kobayashi-Maskawa mechanism~\cite{Cabibbo:1963yz,Kobayashi:1973fv,Christenson:1964fg, Belle:2001qdd, BaBar:2001ags, Belle:2004nch, BaBar:2004gyj, LHCb:2019hro}. In addition, there has also been recent evidence suggesting the existence of $CP$ violation in $\Lambda_b^0$ decay~\cite{LHCb:2024yzj}. However, the observed violations of $CP$ symmetry in these processes are not sufficient to explain the matter-antimatter asymmetry of the universe~\cite{Peskin:2002mm}, implying the existence of  sources of sizable $CP$ violation beyond the SM (BSM)~\cite{He:1999bv, Tandean:2003fr, Salone:2022lpt, He:2022bbs, Goudzovski:2022vbt}.
 Hyperon weak decays have recently been shown to be a promising hunting-ground for $CP$-violating processes. 
In these decays parity violation is described by non-zero values of the decay parameters $\alpha_h$ and $\bar\alpha_h$ for hyperons and anti-hyperons, respectively~\cite{Lee:1957he}.
$CP$ symmetry can be tested experimentally by defining the $CP$-odd asymmetry~\cite{BESIII:2021ypr,BESIII:2020fqg,BESIII:2023sgt,BESIII:2022qax}
\begin{linenomath*}
\begin{equation}
A_{CP}=(\alpha_h+\bar\alpha_h)/(\alpha_h-\bar\alpha_h),
\label{eq:cp_cal}
  \end{equation}
  \end{linenomath*}
where a value significantly deviating from zero would indicate $CP$ violation.
Predictions for the $A_{CP}(\Sigma^+\to p\pi^0)$ asymmetry in the SM vary between $-3.2\times10^{-7}$~\cite{Donoghue:1986hh} and $3.6\times10^{-6}$~\cite{Tandean:2002vy}. Several BSM theories predict an order of magnitude larger values, for example $-3.2\times10^{-5}$ with the Weinberg-Higgs model and $-2.9\times10^{-5}$ with the left-right-symmetric model~\cite{Donoghue:1986hh}.
In 2020, the BESIII collaboration measured this asymmetry for the first time, reporting $A_{CP}(\Sigma^+\to p\pi^0) = 0.004\pm0.037\pm0.010$~\cite{BESIII:2020fqg}, which is consistent with $CP$ conservation. The measurement was based on $1.311\times 10^9$ $J/\psi$ and $4.481\times 10^8$ $\psi(3686)$ events~\cite{BESIII:2016kpv, BESIII:2017tvm}. However, to test the aforementioned theoretical predictions and to search for the potential BSM physics, more precise measurements are needed.

The decay asymmetries that these tests rely upon are experimentally accessible through the angular distribution of the decaying hadron. For a spin-1/2 hyperon decaying into a spin-1/2 baryon and a pseudoscalar meson, the angular distribution is described by
\begin{linenomath*}
\begin{equation}
\mathrm{d}N/\mathrm{d}\Omega = \frac{1}{4\pi}(1 + \alpha_{h} \cdot |\vec{\mathbf{P}}_{h}| \cdot \cos\theta_{P}),
\label{eq:cp_pola}
  \end{equation}
 \end{linenomath*}
where $\vec{\mathbf{P}}_{h}$ is the hyperon polarization, and $\theta_{P}$ is the angle between the polarization direction and the direction of the final state baryon in the hyperon rest frame.  The decay parameter $\alpha_h$ is interesting in its own right. Since it is related to the hyperon matrix element, it gives insight to the SU(3)-flavour structure and the symmetry-breaking patterns of the quark distributions~\cite{Bickerton:2019nyz}. The non-perturbative nature of the strong-decay amplitudes makes this parameter challenging to calculate. Effective field theories~\cite{Borasoy:1998ku} and quark or diquark models~\cite{LeYaouanc:1978ef} have been applied with a varying degree of success. Recent progress in Lattice QCD~\cite{Bickerton:2019nyz}, where calculations of hyperon decay amplitudes are foreseen, opens up the possibility to understand hyperon decays in a more coherent way.

Predictions of the spin properties of heavy multi-strange, charm and beauty baryons rely on precise knowledge of the decay parameters of the light hyperons that they decay into. This is not only true for $CP$ tests of heavy hyperons and baryons~\cite{BESIII:2019odb, Belle:2021crz, Belle:2022uod, LHCb:2021byf}, but also for nuclear and hadron physics, where they are crucial components for understanding the spin structure of hyperons~\cite{Burkardt:1993zh,Mulders:1995dh,Anselmino:2000vs}, the vortex structure~\cite{Becattini:2020ngo}, and the magnetic-field distribution in quark-gluon plasma (QGP)~\cite{Becattini:2016gvu}. Recent expreiments, such as the STAR experiment on $\Lambda$ hyperon polarization in Au+Au collisions~\cite{STAR:2017ckg} and the Belle collaboration's measurement of the transverse polarization of hyperons from quark-antiquark fragmentation~\cite{Belle:2018ttu}, have revealed key insights. However, none of these studies employed the $\Sigma^+$ baryon as a probe. In summary, there are many motivations  for performing  precise measurements of the decay parameters of $\Sigma^{+} \to p \pi^{0}$ and $\bar{\Sigma}^{-} \to \bar{p} \pi^{0}$.

\begin{figure}[hbtp]
\begin{center}
\begin{overpic}[width=0.45\textwidth, trim=20 100 20 20,angle=0]{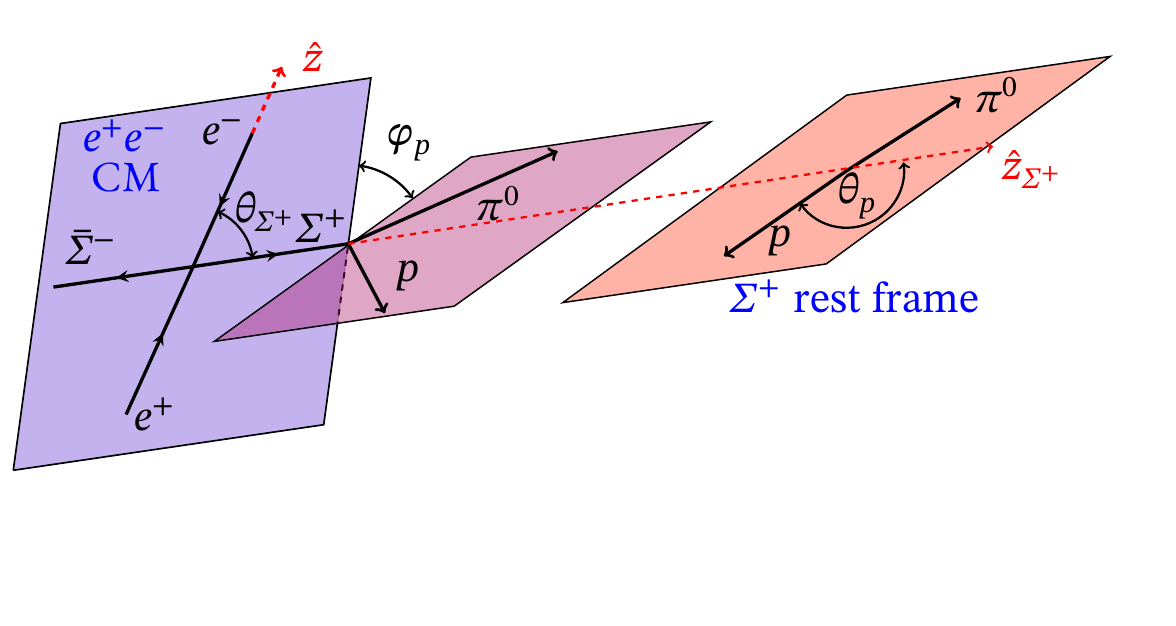}
\end{overpic}
\end{center}
\caption{The helicity frame definition for the decay processes $\Psi \to\Sigma^+\bar\Sigma^-$, $\Sigma^+\to p \pi^0$,
 $\bar\Sigma^-\to \bar{p} \pi^0$ in the center-of-mass (CM) system of the $e^{+} e^{-}$ collision. The helicity angles are defined with respect to the direction of the outgoing hyperon in the CM frame, where $\hat{z}_{\Sigma^+}$ is defined along the $\Sigma^+$ momentum direction in the CM of the $e^{+} e^{-}$ collision.
 }
\label{fig:helix2}
\end{figure}

In the BESIII experiment, the polarized hyperon and anti-hyperon pairs are produced through $e^+e^-$ annihilation, providing a unique environment for studying the production and its decay properties of this system.
The decay parameters and $A_{CP}$ can be measured precisely due to the $\Sigma^+$ and $\bar\Sigma^-$ being in a quantum-entangled state with correlated spin.
The full process $e^+e^-\to \Psi \to\Sigma^+\bar\Sigma^-\to p \pi^0  \bar p \pi^0$ can be described by the helicity angles of the final-state particles, an angular-distribution parameter $\alpha_{\Psi}$, a relative phase $\Delta\Phi$, and two decay parameters~\cite{Faldt:2017kgy, Dubnickova:1992ii, Gakh:2005hh, Faldt:2013gka}.
As seen from the basis vector definitions in Fig.~\ref{fig:helix2}, $\theta_{\Sigma^+}$ is the angle between the $\Sigma^+$ momentum and the electron-beam direction, and $\theta_p$ and $\phi_p$ are the polar and azimuthal angles of the $p$ momentum direction in the $\Sigma^+$ rest frame.
The joint angular distribution ${\cal{W}}({\boldsymbol{\xi}})$~\cite{Faldt:2017kgy, Faldt:2013gka, Faldt:2016qee} of the $e^+e^-\to \Psi \to\Sigma^+\bar\Sigma^-$, $\Sigma^+\to p \pi^0$, and  $\bar\Sigma^-\to \bar p \pi^0$ processes is given by
\begin{linenomath*}
\begin{equation}
\footnotesize
\begin{split}
	{\cal{W}}({\boldsymbol{\xi}})=	 &{\cal{T}}_0({\boldsymbol{\xi}})+{{\alpha_{\Psi}}}{\cal{T}}_5({\boldsymbol{\xi}})\\
	+&{{\alpha_0}{\bar{\alpha}_0}}\left({\cal{T}}_1({\boldsymbol{\xi}})
+\sqrt{1-\alpha_{\Psi}^2}\cos({{\Delta\Phi}}){\cal{T}}_2({\boldsymbol{\xi}})
+{{\alpha_{\Psi}}}{\cal{T}}_6({\boldsymbol{\xi}})\right)\\
+&\sqrt{1-\alpha_{\Psi}^2}\sin({{\Delta\Phi}})
\left({\alpha_0}{\cal{T}}_3({\boldsymbol{\xi}})
+\bar{{\alpha}}_0{\cal{T}}_4({\boldsymbol{\xi}})\right),
\label{eq:anglW}
\end{split}
\end{equation}
  \end{linenomath*}
where $\Psi$ signifies $J/\psi$ or $\psi(3686)$, $\boldsymbol{\xi}$ denotes the polar angle and azimuthal angle of the particle in its final state, ${\cal{T}}_k~(k=0,1,...,6)$ is a set of seven angular functions, and $\alpha_0$ ($\bar{\alpha}_0$) is the decay parameter of $\Sigma^+\to p \pi^0$ ($\bar\Sigma^-\to \bar p \pi^0$). The explicit dependence on the decay parameters $\alpha_0$ and $\bar{\alpha}_0$ enables their simultaneous extraction. Equation~\ref{eq:anglW} consists of three main contributions: one unpolarized part (involving ${\cal{T}}_0$ and ${\cal{T}}_5$) describing the scattering angular distribution, one part (with ${\cal{T}}_1$, ${\cal{T}}_2$ and ${\cal{T}}_6$) describing the spin correlation between the $\Sigma^+$ and the $\bar\Sigma^-$, and finally the spin polarization of the $\Sigma^+$ and $\bar\Sigma^-$ (with ${\cal{T}}_3$ and ${\cal{T}}_4$, respectively).
The spin-correlated part and the spin-polarization part are governed by the psionic form-factor phase $\Delta \Phi$ and depends on the $\Sigma^+$ scattering angle via:
\begin{linenomath*}
\begin{equation}
\vec{\mathbf{P}}_{h}(\cos\theta_{\Sigma^+}) = \frac{\sqrt{1-\alpha_{\Psi}^2}\sin({{\Delta\Phi}})\cos\theta_{\Sigma^+}\sin\theta_{\Sigma^+}}{1+\alpha_{\Psi}\cos^{2}\theta_{\Sigma^+}}.
\label{eq:polar_define}
  \end{equation}
    \end{linenomath*}
A non-zero phase $\Delta\Phi$ means that the $\Sigma^+$ and $\bar\Sigma^-$ hyperons are polarized even if the initial state is unpolarized.

The BESIII experiment has now accumulated $(1.0087\pm0.0044)\times10^{10}$ $J/\psi$ events~\cite{BESIII:2024lks} and $(2.7124\pm0.0143)\times10^{9}$ $\psi(3686)$ events~\cite{BESIII:2021cxx}, which are 7.7 and 6.1 times larger, respectively, than the samples used in the previous measurement~\cite{BESIII:2020fqg}.
This opens unique and excellent opportunity for a new generation of high-precision $CP$ tests, as already demonstrated in several recent papers~\cite{BESIII:2022qax,BESIII:2023sgt,BESIII:2023drj}.
Details about the design and performance of the BESIII detector at the BEPCII collider are given in Ref.~\cite{Ablikim:2009aa}.

The full reaction, including all subsequent decays, is $e^+e^-\to\Psi \rightarrow \Sigma^+\bar\Sigma^-\to p\bar p\pi^0\pi^0\to p\bar p\gamma\gamma\gamma\gamma$.
Selected event candidates are required to have two well-reconstructed charged tracks with net zero charge.
Charged tracks detected in the multi-layer drift chamber (MDC) are required to be within a polar angle ($\theta$) range of $|\rm{cos\theta}|<0.93$, where $\theta$ is defined with respect to the $z$-axis, which is the symmetry axis of the MDC. For each charged track, the distance of closest approach to the interaction point (IP) must be less than 10\,cm along the $z$-axis, and less than 2\,cm in the transverse plane.
Particle identification~(PID) for charged tracks combines measurements of the specific ionization energy loss in the MDC~(d$E$/d$x$) and the flight time in the time-of-flight to form likelihoods $\mathcal{L}(h)~(h=p,K,\pi)$ for each hadron $h$ hypothesis.
Tracks are identified as protons or anti-protons when these hypotheses have the greatest likelihood ($\mathcal{L}(p)>\mathcal{L}(K)$ and $\mathcal{L}(p)>\mathcal{L}(\pi)$).
The distance of closest approaches of the proton and anti-proton tracks to the IP is required to be larger than 0.34\,cm, a distance optimized to suppress background events directly generated from the IP.

Photon candidates are identified using showers in the electromagnetic calorimeter (EMC). The deposited energy of each shower must be more than 25\,MeV in the barrel region ($|\!\cos \theta|< 0.80$) and more than 50\,MeV in the end-cap region ($0.86 <|\!\cos \theta|< 0.92$). To exclude showers originating from the charged tracks, the opening angle subtended by the EMC shower and the position of the closest charged track at the EMC must be greater than $10^\circ$ as measured from the IP. To suppress electronic noise and showers unrelated to the signal event, the difference between the EMC time and the event start-time is required to be within [0, 700]\,ns. To identify the $\pi^0$ candidates, the invariant mass $M_{\gamma\gamma}$ of the two photons has to satisfy $0.115 <M_{\gamma\gamma}< 0.150$\,GeV$/c^2$. In addition, a one-constraint (1C) kinematic fit is performed to constrain $M_{\gamma\gamma}$ to the known $\pi^0$ mass~\cite{ParticleDataGroup:2020ssz}. The $\chi^{2}_{1\rm C}$ of the fit is required to be smaller than 200.
At least one $\pi^0$ candidate per event is selected.  
To further refine the selection, a two-constraint (2C) kinematic fit is performed to the decay $\Psi\to p\pi^0\bar{p}\pi^0$, utilizing four-momentum conservation in the decay and the known $\pi^0$ mass from the $\Sigma^+$ decay, while the three-momentum of the $\pi^0$ from the $\bar\Sigma^-$ is not required but treated as an  unknown set of variables. This reduces the contamination from secondary showers induced by anti-protons annihilating in the EMC. If there is more than one $\pi^0$ candidate, that one with the smallest $\chi^{2}_{2\rm C}$ is retained. A requirement on the quality of the 2C kinematic fit of $\chi^2_{\rm 2C} < 30$ is imposed to select the signal candidates.

\begin{figure}[hbtp]
\begin{center}
\begin{overpic}[width=0.4\textwidth, trim=50 80 60 50,angle=0]{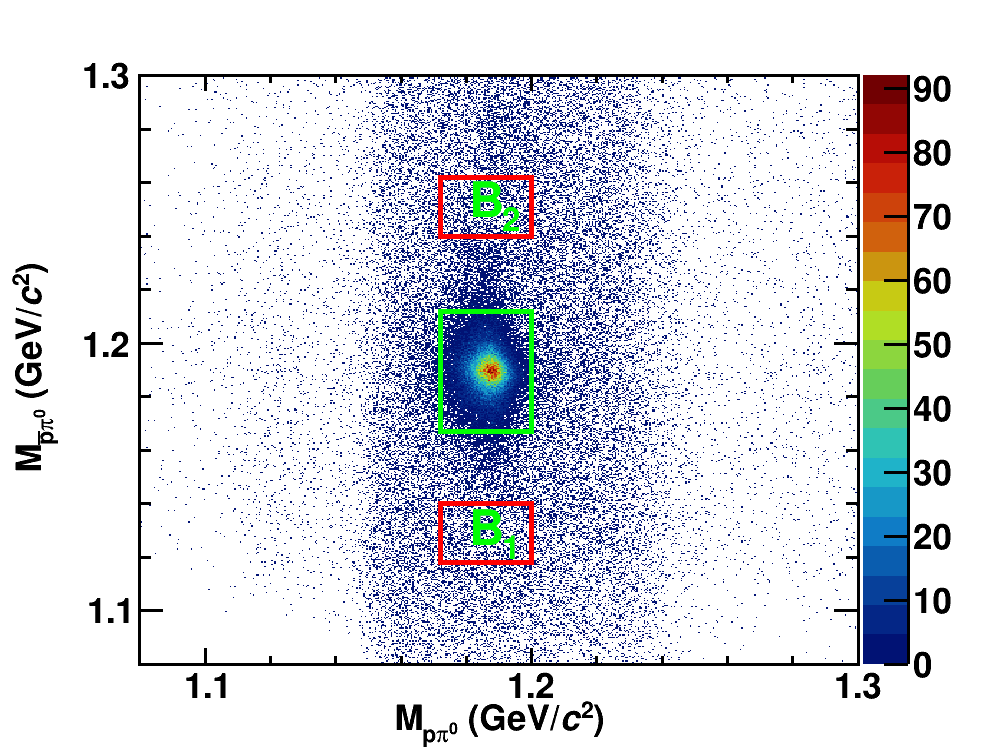}
\end{overpic}
\end{center}
 \caption{Distribution of $M_{\bar p \pi^{0}}$ versus $M_{p \pi^{0}}$ for the $\psi(3686)$ data
sample. The variation in event density distribution is caused by the kinematic fit. The green box denotes the signal region, and the red ones indicate the sideband regions.}
\label{fig:sideband}
\end{figure}

The inclusive Monte Carlo (MC) samples of $10^{10}$ $J/\psi$ events and $2.7\times10^9$ $\psi(3686)$ events are used to investigate possible sources of contamination.  MC events passing the event selection are examined with TopoAna, a generic event-type analysis tool~\cite{ref:topo}. All particle decays are modeled with {\sc evtgen}~\cite{ref:evtgen} using branching fractions either taken from the Particle Data Group (PDG)~\cite{ParticleDataGroup:2020ssz}, when available, or otherwise estimated by MC simulations based on {\sc lundcharm}~\cite{ref:lundcharm}.
The main peaking background contributions are $\Psi\to\gamma\Sigma^+\bar\Sigma^-$ and $\Psi\to\gamma\eta_{c}$ ($\eta_{c}\to\Sigma^+\bar\Sigma^-$), with fractions estimated to be 0.1\%.  The main source of non-peaking background are the decays $\Psi\to\Delta^+\bar\Delta^-\to p\pi^0\bar p\pi^0$ and $\Psi\to p\pi^0\bar p\pi^0$, which contribute  0.9\% and 0.7\% of the total data samples for $J/\psi$ and $\psi(3686)$, respectively.
Given the low level of peaking background, the sideband method is suitable for performing the background subtraction.
Figure~\ref{fig:sideband} shows the distribution of the invariant masses $M_{\bar p\pi^0}$ versus $M_{p\pi^0}$ from $\psi(3686)$ decays. The signal region in the green rectangle is defined as $1.167 < M_{\bar p\pi^0} < 1.212$\,GeV/$c^2$ and $1.172 < M_{p\pi^0} < 1.200$\,GeV/$c^2$. To determine the background contributions in the signal region, the sideband regions in the red rectangles are defined as $1.118 < M_{\bar p\pi^0} < 1.140$\,GeV/$c^2$, $1.240 < M_{\bar p\pi^0} < $1.262\,GeV/$c^2$, and $1.172 < M_{p\pi^0} < 1.200$\,GeV/$c^2$. The mass windows are approximately $\pm3\sigma$ away from the known $\bar\Sigma^-$ mass~\cite{ParticleDataGroup:2020ssz}, with $\sigma$ being the $M_{\bar p\pi^0}$ mass resolution. The background yields are estimated to be $f \times (B_{1}+B_{2})$, where $B_{1}$ and $B_{2}$ denote the number of background events in the rectangles, and $f=1.06$ denotes the ratio of the background events between the signal and sideband regions determined by fitting the distribution of $M_{\bar p\pi^0}$. Here, the signal and background candidates are described by the signal MC shape and second-order polynomial function, respectively.

\begin{figure}[hbtp]
\begin{center}
\begin{overpic}[width=0.4\textwidth, trim=40 50 50 20,angle=0]{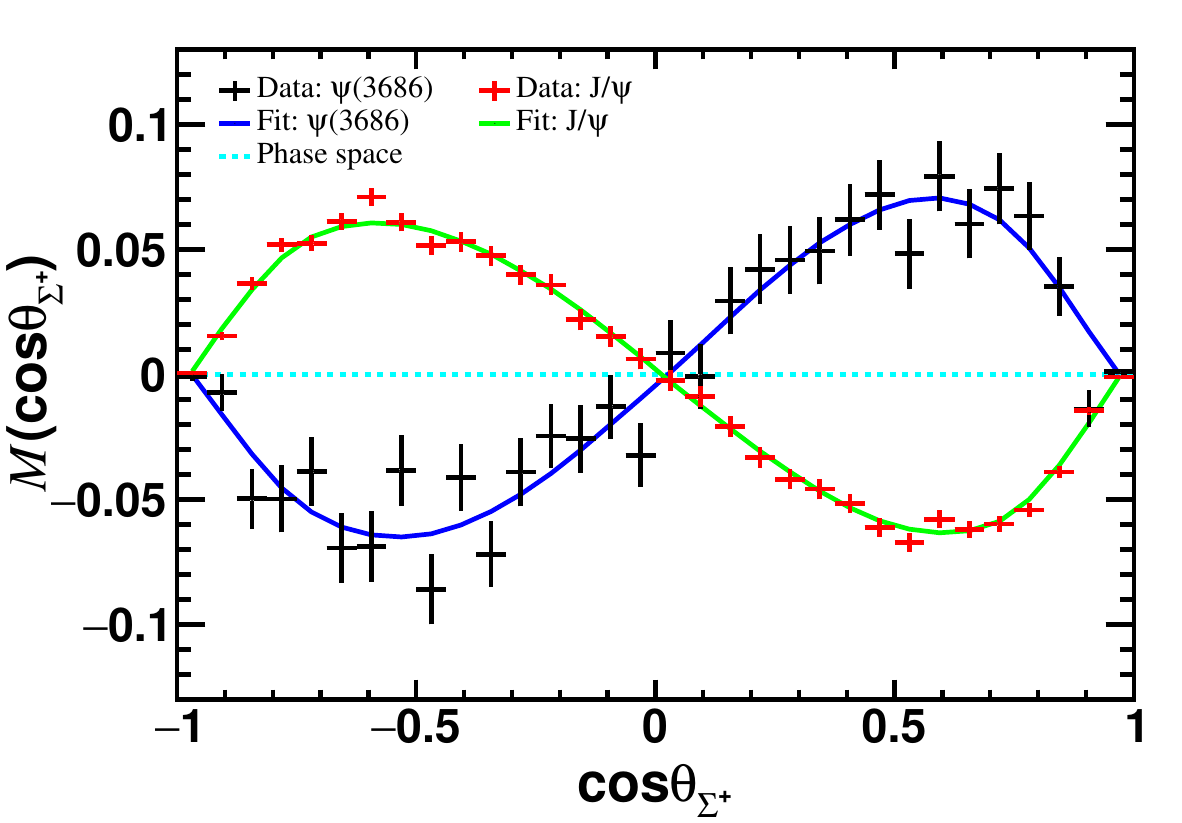}
\end{overpic}
\end{center}
 \caption{The moment $M(\cos\theta_{\Sigma^{+}})$ is a function
  of $\cos\theta_{\Sigma^{+}}$ for
  $\Psi\to\Sigma^{+}\bar{\Sigma}^{-}\to p\pi^0\bar p\pi^0$. The black and red data points with error bars are obtained independently via Eq.~\ref{fig:moment} for $\psi(3686)$ and $J/\psi$ decays. The blue and green solid lines are drawn with the measured parameters, and the cyan dashed line represents the distribution without polarization uniformly distributed in  phase space.
 }
\label{fig:polarization}
\end{figure}

An unbinned maximum likelihood fit is performed with the five measured angles ${\boldsymbol{\xi}}$ as input. The fit is executed on the $J/\psi$ and $\psi(3686)$ samples simultaneously and the free parameters are $\alpha_{J/\psi}$, $\alpha_{\psi(3686)}$, $\Delta\Phi_{J/\psi}$, $\Delta\Phi_{\psi(3686)}$, $\alpha_{0}$, and $\bar\alpha_{0}$. 
Following the approach in Ref.~\cite{BESIII:2022qax}, the reconstruction efficiency is taken into account in a model-independent way and the background contribution is included with the scale factor $f$. The numerical fit results are summarized in Table~\ref{sys-tot1}, where the derived values of $\left< \alpha_{\rm 0} \right>$ and $A_{CP}$ are also included. The relative phases between the $\Psi$ electric and magnetic form factors are determined to
be $\Delta\Phi_{J/\psi} = -0.2744 \pm 0.0033 \pm 0.0010$~rad and $\Delta\Phi_{\psi(3686)} = 0.427 \pm 0.022 \pm 0.003$~rad, respectively. This difference in sign indicates that the $\Sigma^+$ is polarized in such a way that its spins are aligned in opposite directions when produced via $J/\psi$ decay compared to $\psi(3686)$ decay. This behavior was observed in an earlier BESIII study~\cite{BESIII:2020fqg, BESIII:2023sgt} and is now confirmed with improved precision.
The polarization can be studied graphically by defining the polarization estimator $M(\cos\theta_{\Sigma^+})$:
\begin{linenomath*} 
\begin{equation}
M(\cos\theta_{\Sigma^{+}}) = \frac{m}{N}\sum_i^{N(\cos\theta_{\Sigma^{+}})}(\sin\theta_{p}^{i} \sin\phi_{p}^{i} - \sin\theta^{i}_{\bar{p}} \sin\phi_{\bar{p}}^{i}),
\label{fig:moment}
\end{equation}
 \end{linenomath*}
 where $m=32$ is the number of bins, $N$ is the total number of events in the data sample, and $N(\cos\theta_{\Sigma^+})$ is the number of events in the $\cos\theta_{\Sigma^+}$ bin. 
Figure~\ref{fig:polarization} shows the $\Sigma^+$ polarization distributions where the opposite polarization directions for $J/\psi$ and $\psi(3686)$ decays are evident.

\begin{table}[hbtp]
 \centering
 \footnotesize
\caption{The decay parameters of $\Psi\rightarrow\Sigma^+\bar\Sigma^-\rightarrow p\pi^0\bar p \pi^0$ where the first uncertainty is statistical and the second systematic.}
\label{sys-tot1}
\begin{tabular}{lccc}
\hline \hline 
Parameter&This Letter&Previous result~\cite{BESIII:2020fqg}\\\hline
$\alpha_{J/\psi}$                                       &$-0.5047 \pm 0.0018\pm0.0010$ &$-0.508\pm0.006\pm0.004$\\
$\Delta\Phi_{J/\psi}$                                 &$-0.2744 \pm 0.0033\pm0.0010$ &$-0.270\pm0.012\pm0.009$\\
$\alpha_{0}$                                             &$-0.975\pm0.011 \pm0.002~~~~$ &$-0.998\pm0.037\pm0.009 $\\
$\bar\alpha_{0}$                                       &$~0.999\pm0.011\pm0.004~~~$ &$~~0.990\pm0.037\pm0.011 $\\
$\alpha_{\psi(3686)}$                                &$~~0.7133 \pm 0.0094\pm0.0065$ &$~~0.682 \pm 0.030\pm0.011$\\
$\Delta\Phi_{\psi(3686)}$                          &$~0.427 \pm 0.022\pm 0.003~~~$ &$~~0.379 \pm 0.070\pm0.014$ \\\hline
$\it{\left< \alpha_{\rm 0}\right>}$                &$-0.9869\pm0.0011\pm0.0016$ &$-0.994\pm0.004\pm0.002$\\
$A_{CP}$                                                  &$-0.0118\pm0.0083\pm0.0028$ &$~~0.004\pm0.037\pm0.010$\\
 \hline \hline
\end{tabular}
\end{table}

\begin{figure}[hbtp]
\begin{center}
\begin{overpic}[width=0.4\textwidth, trim=35 50 45 20,angle=0]{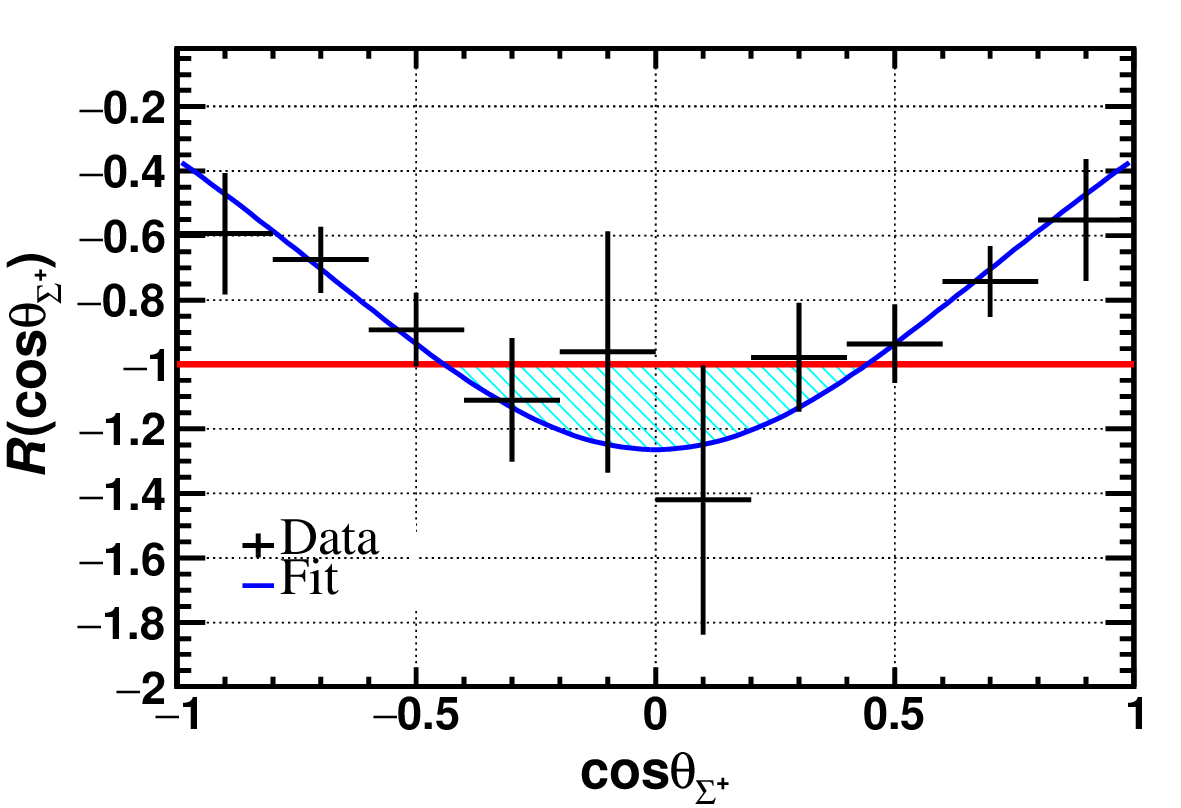}
\end{overpic}
\end{center}
 \caption{The angular distribution of the ratio of polarization between $J/\psi$ and $\psi(3686)$ decays. 
The black dots with error bars are the data, the blue solid line is the fitting results, 
The red line represents a ratio of $-1$. Comparing the $\Sigma^+$ from the $J/\psi$ decay, the range of $\cos\theta_{\Sigma^+}$ in the shaded area (perpendicular to the electron beam direction) indicates the enhanced polarization of the $\Sigma^+$ in  $\psi(3686)$ decays.
 }
\label{fig:polarization_r}
\end{figure}

To investigate further the difference between $J/\psi$ and $\psi(3686)$ decays, the quantity $R(\cos\theta_{\Sigma^+})$ is defined as the ratio  of the polarization values ($\vec{\mathbf{P}}_{h}$), measured according to Eq.~\ref{eq:polar_define}, between $J/\psi$$\to\Sigma^{+}\bar{\Sigma}^{-}$ and $\psi(3686)$$\to\Sigma^{+}\bar{\Sigma}^{-}$ in each bin of  $\cos\theta_{\Sigma^+}$~\cite{Faldt:2017kgy}. In Fig.~\ref{fig:polarization_r}, $R(\cos\theta_{\Sigma^+})$ is shown as a function of the $\Sigma^+$ scattering angle. $R(\cos\theta_{\Sigma^+})$ is negative over the full angular range, indicating that the sign of the polarization is opposite between the two decays in a given angular bin, consistent with the previous measurement of $\Psi \rightarrow \Sigma^0 \bar{\Sigma}^{0}$~\cite{BESIII:2024nif}.
The larger sample size allows the magnitude of $R(\cos\theta_{\Sigma^+})$, which reflects the relationship between polarization differences in $J/\psi$ and $\psi(3686)$ decays and the scattering angle, to be determined with interesting precision. It is found that the magnitude of $R(\cos\theta_{\Sigma^+})$  is smallest near $\cos\theta_{\Sigma^+}=\pm1$, which corresponds to $\Sigma^+$ hyperons moving parallel ($+$) or antiparallel ($-$) to the positron beam. This is expected since at these angles, the production plane is not well-defined, and the polarization must therefore be zero.  The magnitude of $R(\cos\theta_{\Sigma^+})$ increases at large angles, and reaches a maximum in the direction perpendicular to the beam axis.

The systematic uncertainties are summarized in Table~\ref{tot_parameter1} where the totals are obtained from adding the individual components in quadrature.

 \begin{table}[hbtp]
 \centering
  \footnotesize
\caption{The uncertainties ($\times10^{-3}$) of the parameters $\alpha_{J/\psi}$, $\Delta\Phi_{J/\psi}$, $\alpha_{0}$, $\bar{\alpha}_{0}$, $\alpha_{\psi(3686)}$, and $\Delta\Phi_{\psi(3686)}$. A dash ($-$) indicates that the uncertainties are negligible.}
\label{tot_parameter1}
\begin{tabular}{lcccccc}
\hline \hline 
Source &$\alpha_{J/\psi}$ &$\Delta\Phi_{J/\psi}$&$\alpha_{0}$&$\bar{\alpha}_{0}$&$\alpha_{\psi(3686)}$ &$\Delta\Phi_{\psi(3686)}$\\\hline 
Efficiency correction       &$0.1$    &$0.3$   &$0.7$    &$0.9$    &$0.2$  &$0.3$ \\ 
Kinematic fit                  &$0.6$       &$-$       &$-$        &$-$        &$1.9$  &$2.1$ \\ 
Signal mass window       &$0.5$   &$0.8$    &$1.5$    &$1.6$    &$5.6$  &$2.0$\\
Background                    &$0.3$   &$0.3$    &$-$        &$-$        &$2.4$  &$-$\\
Fitting method                &$0.6$   &$0.3$    &$0.6$    &$3.6$    &$1.4$   &$0.1$\\
\hline
 Total                              &$1.0$    &$1.0$    &$1.8$    &$4.0$    &$6.5$  &$3.0$\\
 \hline \hline
\end{tabular}
\end{table}

The ratios of differences in efficiency between data and MC simulation for charged-particle tracking, PID, and $\pi^0$ reconstruction are studied using the  control samples $J/\psi\to p\bar p\pi^+\pi^-$ and $J/\psi\to\Sigma^+\bar\Sigma^-\to p\pi^0 \bar p\pi^0$. The calculated ratios are then used as correction factors. Applying Gaussian sampling to the correction factors, and correcting samples with different factors, the phase-space MC sample is corrected 100 times and is used to determine the fit results. The uncertainty of the aforementioned ratios are taken as the width of the distribution of fit results. The average difference compared with the baseline result is taken as a systematic uncertainty.
The uncertainty from the kinematic fit is estimated by varying the selection requirement of  $\chi_{2\rm C}^{2}\le30$ from 20 to 40.  
The uncertainty due to the requirement of signal mass window is determined by changing the window range.
The uncertainty associated with the background subtraction is evaluated by varying the sideband window within a range of 4\,MeV/$c^2$.
We use the same criteria as for $\chi_{2\rm C}^{2}\le30$ to quantify variations that cannot be explained by statistical fluctuations. The largest difference with respect to the baseline result is taken as a systematic uncertainty.
To evaluate the reliability of parameter estimation with the maximum likelihood fit, a set of 100 MC samples is produced with  {\sc evtgen} and propagated through the BESIII detector. We use the angular distributions according to Eq.~\ref{eq:anglW}, with the parameter values obtained from the data analysis, and inspect the difference between input and output distributions. The difference is taken as the systematic uncertainty of the fitting method.

In summary,  a five-dimensional angular analysis is simultaneously performed to the processes $J/\psi \to\Sigma^+\bar\Sigma^-$ and $\psi(3686) \to\Sigma^+\bar\Sigma^-$ selected from $(1.0087\pm0.0044)\times10^{10}$ $J/\psi$ events and $(2.7124\pm0.0143)\times10^{9}$ $\psi(3686)$ events collected at the BESIII detector. Exploiting the quantum-entangled $\Sigma^+\bar\Sigma^-$ system and the self-analysing $\Sigma$ decays, the measured precision of the parameters related to $\Sigma$ pair production are improved by factors of three for $\alpha_{\psi(3686)}$, $\alpha_{J/\psi}$ and $\Delta\Phi_{\psi(3686)}$, and four for $\Delta\Phi_{J/\psi}$. The precision of the decay parameters $\alpha_{0}$ and $\bar\alpha_{0}$ are improved by a factor of three and are consistent with the PDG average. The $CP$ asymmetry, $A_{CP}=-0.0118\pm0.0083\pm0.0028$, is consistent with $CP$ conservation within 1.3$\sigma$. The average value $\it{\left< \alpha_{\rm 0}\right>}=$ $-0.9869\pm0.0011\pm0.0016$ is the most precise measurement in all baryon decays, and is a crucial input for precise decay parameter measurements and $CP$ violation searches in the charm and beauty baryons decays to $\Sigma^+$. Furthermore, it provides input for the observable measurements related to QGP in high-energy nuclear collisions. A detailed comparison of hyperon polarization with sign flip between $J/\psi$ and $\psi(3686)$ decays are performed for the first time.
Our measurements are expected to remain the most precise for $\Sigma$ hyperons over the next decade, until results become available from future higher-luminosity machines~\cite{Salone:2022lpt,Goudzovski:2022vbt,Achasov:2023gey}.

The BESIII Collaboration thanks the staff of BEPCII and the IHEP computing center for their strong support. This work is supported in part by National Key R\&D Program of China under Contracts Nos. 2023YFA1606000, 2020YFA0406300, 2020YFA0406400; National Natural Science Foundation of China (NSFC) under Contracts Nos. 12375070, 11635010, 11735014, 11935015, 11935016, 11935018, 12025502, 12035009, 12035013, 12061131003, 12192260, 12192261, 12192262, 12192263, 12192264, 12192265, 12221005, 12225509, 12235017, 12361141819; the Chinese Academy of Sciences (CAS) Large-Scale Scientific Facility Program; the CAS Center for Excellence in Particle Physics (CCEPP); Joint Large-Scale Scientific Facility Funds of the NSFC and CAS under Contract No. U1832207; CAS under Contract No. YSBR-101; 100 Talents Program of CAS; The Institute of Nuclear and Particle Physics (INPAC) and Shanghai Key Laboratory for Particle Physics and Cosmology; Agencia Nacional de Investigación y Desarrollo de Chile (ANID), Chile under Contract No. ANID PIA/APOYO AFB230003; German Research Foundation DFG under Contract No. FOR5327; Istituto Nazionale di Fisica Nucleare, Italy; Knut and Alice Wallenberg Foundation under Contracts Nos. 2021.0174, 2021.0299; Ministry of Development of Turkey under Contract No. DPT2006K-120470; National Research Foundation of Korea under Contract No. NRF-2022R1A2C1092335; National Science and Technology fund of Mongolia; National Science Research and Innovation Fund (NSRF) via the Program Management Unit for Human Resources \& Institutional Development, Research and Innovation of Thailand under Contract No. B50G670107; Polish National Science Centre under Contract No. 2019/35/O/ST2/02907; Swedish Research Council under Contract No. 2019.04595; The Swedish Foundation for International Cooperation in Research and Higher Education under Contract No. CH2018-7756; U. S. Department of Energy under Contract No. DE-FG02-05ER41374.

\end{document}

%% file: def-com.tex
\newcommand{\bfg}{\begin{figure}}
\newcommand{\efg}{\end{figure}}
\newcommand{\bitm}{\begin{itemize}}
\newcommand{\eitm}{\end{itemize}}
\newcommand{\bnum}{\begin{enumerate}}
\newcommand{\enum}{\end{enumerate}}
\newcommand{\btbl}{\begin{table}}
\newcommand{\etbl}{\end{table}}
\newcommand{\btbu}{\begin{tabular}}
\newcommand{\etbu}{\end{tabular}}
\newcommand{\bcl}{\begin{center}}
\newcommand{\ecl}{\end{center}}

\newcommand{\beq}{\begin{equation}}
\newcommand{\eeq}{\end{equation}}
\newcommand{\beqr}{\begin{eqnarray}}
\newcommand{\eeqr}{\end{eqnarray}}

\newcommand{\blue}{\color{blue}}